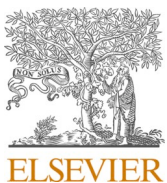
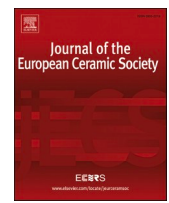

# Quantifying Li-content for compositional tailoring of lithium ferrite ceramics

C. Granados-Miralles [a,*], A. Serrano [a], P. Prieto [b], J. Guzmán-Mínguez [a], J.E. Prieto [c], A.M. Friedel [d,e], E. García-Martín [b,c], J.F. Fernández [a], A. Quesada [a]

[a] *Instituto de Cerámica y Vidrio, CSIC, ES-28049 Madrid, Spain*
[b] *Dpto. de Física Aplicada, Universidad Autónoma de Madrid, Madrid ES-28049, Spain*
[c] *Instituto de Química Física 'Rocasolano', CSIC, Madrid ES-28006, Spain*
[d] *Institut Jean Lamour, UMR CNRS 7198 and Université de Lorraine, FR-54000 Nancy, France*
[e] *Fachbereich Physik and Landesforschungszentrum OPTIMAS, Technische Universität Kaiserslautern, DE-67663 Kaiserslautern, Germany*

A B S T R A C T

Owing to their multiple applications, lithium ferrites are relevant materials for several emerging technologies. For instance, $LiFeO_2$ has been spotted as an alternative cathode material in Li-ion batteries, while $LiFe_5O_8$ is the lowest damping ferrite, holding promise in the field of spintronics. The Li-content in lithium ferrites has been shown to greatly affect the physical properties, and in turn, the performance of functional devices based on these materials. Despite this, lithium content is rarely accurately quantified, as a result of the low number of electrons in Li hindering its identification by means of routine materials characterization methods. In the present work, magnetic lithium ferrite powders with Li:Fe ratios of 1:1, 1:3 and 1:5 have been synthesized, successfully obtaining phase-pure materials ($LiFeO_2$ and $LiFe_5O_8$), as well as a controlled mixture of both phases. The powders have been compacted and subsequently sintered by thermal treatment ($T_{max} = 1100$ °C) to fabricate dense pellets which preserve the original Li:Fe ratios. Li-content on both powders and pellets has been determined by two independent methods: (i) Rutherford backscattering spectroscopy combined with nuclear reaction analysis and (ii) Rietveld analysis of powder X-ray diffraction data. With good agreement between both techniques, it has been confirmed that the Li:Fe ratios employed in the synthesis are maintained in the sintered ceramics. The same conclusion is drawn from spatially-resolved confocal Raman microscopy experiments on regions of a few microns. Field emission scanning electron microscopy has evidenced the substantial grain growth taking place during the sintering process – mean particle sizes rise from $\approx 600$ nm in the powders up to 3.8(6) μm for dense $LiFeO_2$ and 10(2) μm for $LiFe_5O_8$ ceramics. Additionally, microstructural analysis has revealed trapped pores inside the grains of the sintered ceramics, suggesting that grain boundary mobility is governed by surface diffusion. Vibrating sample magnetometry on the ceramic samples has confirmed the expected soft ferrimagnetic behavior of $LiFe_5O_8$ (with $M_s = 61.5(1)$ Am$^2$/kg) and the paramagnetic character of $LiFeO_2$ at room temperature. A density of 92.7(6)% is measured for the ceramics, ensuring the mechanical integrity required for both their direct utilization in bulk shape and their use as targets for thin-film deposition.

## 1. Introduction

Lithium ferrites (i.e, $LiFeO_2$, $LiFe_5O_8$) are materials of great interest due to their multiple and varied properties and applications. More specifically, the rock-salt lithium ferrite, $LiFeO_2$, is an antiferromagnetic (AFM) material below $\approx 90$ K and paramagnetic at room temperature (RT) [1,2]. $LiFeO_2$ has drawn great attention as an alternative to $LiCoO_2$, the most widely used cathode material in commercial Li-ion batteries. With a very similar atomic structure, $LiFeO_2$ has been proposed as a candidate for substituting the Co-based ferrite owing to the greater abundance, lower price and non-toxicity of Fe compared to that of Co. [1,3–5]. $LiFeO_2$ is also a good chemical sorbent for $CO_2$, which is used as a strategy to reduce the amount of $CO_2$ released to the atmosphere [6,7], and it has also been proposed as electrocatalyst material for a sustainable $NH_3$ production through reduction of $N_2$ [8]. Additionally, the unusual optical transitions recently reported for $LiFeO_2$ open the door for using this material for various spintronic and photo-catalysis applications [9].

The spinel lithium ferrite, LiFe$_5$O$_8$, is a soft ferrimagnetic (FiM) material with great technological significance. Spinel ferrites in general are widely used in microwave devices (e.g. isolators, circulators, phase shifters, absorbers) [10,11]. Compared to other soft spinel ferrites, LiFe$_5$O$_8$ stands out with the highest Curie temperature ($T_C \approx 950$ K) [10,12] and the lowest losses at high microwave frequencies [13,14]. Some years ago, a Gilbert damping parameter of $2.1 \times 10^{-3}$ was measured for a LiFe$_5$O$_8$ single-crystal [13], and a more recent study has reported a value as low as $1.3 \times 10^{-3}$ for an epitaxial LiFe$_5$O$_8$ thin film, drawing attention to this material as a promising candidate for spintronic applications. Other applications of LiFe$_5$O$_8$ include gas sensing or anode material for Li-ion batteries [15,16].

A good number of studies on both lithium ferrite phases have been cited above, in which single-crystals, thin films or powder samples are studied based on a plethora of characterization techniques, e.g. X-ray and neutron diffraction, Raman, infrared and Mössbauer spectroscopy, scanning and transmission electron microscopy, etc., as well as various physical property measurement, e.g., charge-discharge voltage profiles, ferromagnetic resonance, magnetic hysteresis, Néel/Curie temperature determination, dielectric properties, etc. Generally speaking, an appropriate and accurate quantification of the Li content is often lacking. The reduced number of electrons in lithium makes it difficult to quantify based on the characterization techniques routinely employed to determine the elemental composition of materials [17]. However, lithium quantification is rather critical given that the Li-content has been demonstrated to have a great influence over some physical properties [18], therefore conditioning the performance of functional devices based on lithium ferrites. Moreover, Li has been seen to have a tendency to escape through evaporation promoted by the elevated temperatures generally employed in the material preparation [14].

In this work, lithium ferrite powders with different Li:Fe ratios have been synthesized. The powders have subsequently conformed into pellets and sintered following a traditional ceramic processing method. Elemental composition of the samples has been extracted from Rutherford backscattering spectroscopy combined with nuclear reaction analysis (RBS-NRA), Rietveld analysis of powder X-ray diffraction (PXRD) data has yielded quantitative phase and elemental compositions as well as unit cell dimensions for the different phases present in each sample, while the samples homogeneity has been investigated by confocal Raman microscopy (CRM). All mentioned techniques demonstrate with good agreement that the phase composition of the powders is essentially maintained on the sintered ceramics, which is crucial in terms of the material functionality. Microstructural analysis has shed light on the processes that occur during sintering of the ceramics, evidencing the role of the Li-content of the starting powders. The magnetic properties of the ceramics have been investigated through vibrating sample magnetometry (VSM) measurements at RT.

## 2. Experimental methods

### 2.1. Sample preparation

Appropriate amounts of Li$_2$CO$_3$ and α-Fe$_2$O$_3$ ($\geq$ 99% and $\geq$ 96%, Sigma-Aldrich) were mixed to attain Li:Fe ratios of 1:1, 1:3 and 1:5. The dry powders were mixed using a LabRAM resonant acoustic mixer from Resodyn (80 g, 1 min at $\approx 61.5$ Hz, max. acceleration = 63 G's). The mixture was thermally treated in air atmosphere (4 h at 900 °C) in a laboratory furnace. Synthesis heating profiles may be found in Fig. S1a in the Supporting Information (SI). The synthesis treatment was optimized based on previous experiments (not shown here) so that a cubic phase with as low unit cell parameter as possible is obtained for the 1:5 Li:Fe ratio mixture (LiFe$_5$O$_8$ a $\approx 8.33$ Å) [13,19,20], given that too low temperature is reported to yield hematite (hexagonal/rhombohedral) and too high temperature have been seen to produce magnetite (cubic, a $\approx 8.39$ Å) [21]. The expected reactions for each Li:Fe ratio are: [22,23].

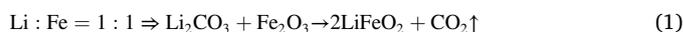

Li : Fe = 1 : 1 ⇒ Li$_2$CO$_3$ + Fe$_2$O$_3$ → 2LiFeO$_2$ + CO$_2$↑   (1)

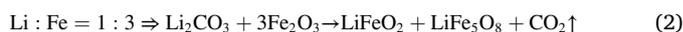

Li : Fe = 1 : 3 ⇒ Li$_2$CO$_3$ + 3Fe$_2$O$_3$ → LiFeO$_2$ + LiFe$_5$O$_8$ + CO$_2$↑   (2)

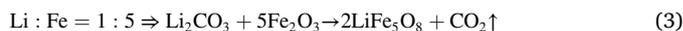

Li : Fe = 1 : 5 ⇒ Li$_2$CO$_3$ + 5Fe$_2$O$_3$ → 2LiFe$_5$O$_8$ + CO$_2$↑   (3)

The obtained powders were gently grinded in an agate mortar, yielding to three different powder samples, which will be subsequently referred to as <Li-Fe 1–1 powd> , <Li-Fe 1–3 powd> and <Li-Fe 1–5 powd> .

The synthesized powders were pressed into cylindrical pellets (ø=1 cm, 0.75 g powders, 6 wt% lubricant) using a stainless-steel die of appropriate dimensions and a manual press from Tonindustrie (5 min, 125 MPa). The pellets were then heated in air atmosphere (4 h at 1100 °C) in the same furnace used for the synthesis, now following a 3-step thermal cycle previously reported to yield high-density LiFe$_5$O$_8$ pieces [24]. (see Fig. S1b for sintering heating profiles). The obtained ceramic samples are given the following IDs: <Li-Fe 1–1 ceram> , <Li-Fe 1–3 ceram> and <Li-Fe 1–5 ceram> .

### 2.2. Sample characterization

The elemental composition of the samples (Li, Fe, O) was evaluated by means of ion beam analysis techniques. RBS combined with NRA was carried out at the 5 MV tandem accelerator at CMAM using H$^+$ at 3.0 MeV [25]. A silicon barrier detector placed at a scattering angle of 150° was used to measure the backscattering yield and the sample position was controlled with a three-axis goniometer. For the measurement, the pellets were grinded to powders using 800 grit dry SiC sandpaper and the obtained powders were fixed on a carbon adhesive tape. The powder samples were directly mounted on the carbon tape. The experimental conditions for the RBS-NRA measurements were specifically chosen to detect and quantify the Li present in the sample. In particular, non-invasive characterization of the lithium content was performed using the [7]Li(p,α)$^4$He nuclear reaction, which cross-section has a broad maximum at a proton energy of 3 MeV. This nuclear reaction is considered the most suitable for lithium quantification since the signal intensity is proportional to the amount of the naturally occurring [7]Li isotope while yielding a high signal-to-noise ratio [26]. The in-depth quantification and distribution of Li and all the other elements in the samples were determined with the SIMNRA simulation software package [27].

PXRD data were collected in the 2θ-range 14–78° using a Bruker D8 Advance X-Ray diffractometer (ω-2θ diffraction scans) equipped with a Lynx Eye detector and a Cu anode (K$_{α1}$(Cu)= 1.5406 Å) operated at 40 kV and 30 mA. The powder samples were measured as-synthesized, while the ceramics were ground into powders prior to the PXRD measurement. Rietveld analysis of the PXRD data was performed with the software *FullProf* [28], using a Thompson-Cox-Hastings pseudo-Voigt function to describe the peak-profile [29]. The crystallographic phases included in the Rietveld models were α-LiFeO$_2$ (rock-salt structure, $Fm\bar{3}m$ (225), a= 4.158(1) Å) [17] and α-LiFe$_5$O$_8$ (spinel structure, $P4_332$ (212), a= 8.3339(1) Å) [30]. Find a detailed description of the phases in Tables S1 and S2, respectively. Notably, LiFe$_5$O$_8$ crystallizes in the ordered α-LiFe$_5$O$_8$ phase rather than the high-symmetry β-LiFe$_5$O$_8$ phase ($Fd\bar{3}m$) (see Fig. S3). A NIST standard (LaB$_6$ SRM® 660b) [31] was measured in the same experimental conditions as the samples in order to estimate the instrumental-contribution to the peak-broadening and a Lorentzian isotropic size parameter was refined to describe the sample-contribution to the broadening. Refinements of the Rietveld models yielded quantitative information on the samples (incl. phase composition, elemental analysis for the individual phases).

Phases lacking long-range order are not discernible based on diffraction data. Therefore, the samples were further characterized by Raman spectroscopy to reveal possible non-crystalline phases. In particular, the samples were investigated using a confocal Raman system (WITec ALPHA 300RA) in which the spectrometer is coupled with an optical microscope, allowing recording spatially-resolved Raman data (2D-Raman maps). The measurements were performed at RT with a linearly p-polarized Nd:YAG laser (532 nm) and a 100x objective lens (NA=0.95). A low laser excitation power (0.5 mW) was used to avoid sample damage and/or overheating. Representative Raman spectra were obtained from averaging single spectra recorded every 200 nm (exposure time = 2 s) throughout an area of at least $5 \times 5$ μm$^2$. Several regions of powdered samples were measured in order to analyze their homogeneity. Raman results were analyzed by using WITec Project Plus Software. In all cases, Raman spectra were normalized to the highest intensity vibration mode. A fraction of the pellets was ground into powders with an agate mortar, while the powder samples were measured as-synthesized.

The density of the ceramics was determined by the Archimedes' method in distilled water at 25 °C. Theoretical densities were calculated for all samples using the refined phase compositions and unit cell dimensions extracted from Rietveld analysis of the PXRD data. Relative densities were calculated dividing measured (Archimedes) by calculated (theoretical) density values.

The morphology and microstructure of the samples were investigated based on secondary electron images of FE-SEM using a Hitachi S-4700 microscope. Fresh fractured surfaces were imaged for the ceramic samples. The powders were measured directly. Grain size distributions and average grain sizes for both powders and pellets were extracted from the FE-SEM images using the software ImageJ [32].

A hot-stage microscope (HSM) from Hesse Instruments with an EMI image analysis and data processing software allowed extracting dilatometric curves for <Li-Fe 1–1 powd> and <Li-Fe 1–5 powd>, which served to study the sintering kinetics of LiFeO$_2$ and LiFe$_5$O$_8$, respectively. For HSM measurements, the sample was placed on an alumina substrate and heated up to 1500 °C with a heating rate of 10 °C/min.

RT magnetic hysteresis of the dense ceramics was investigated using a vibrating sample magnetometer (VSM) from Microsense (model EZ 7). The cylindrical pellets were mounted on a quartz rod with a quartz disk on one end onto which the samples were fixed with Teflon tape. VSM measurements were performed at RT in a field range of up to 2.1 T. Saturation magnetization, $M_s$, and coercive field, $H_c$, values are extracted from the measured hysteresis loops; $M_s$ being the magnetization value at the maximum applied field (i.e., $M_s = M_{(2.1\,T)}$) and $H_c$ the $H$-field value at $M = 0$.

## 3. Results and discussion

### 3.1. Elemental and phase composition. Structural characterization

Three compositions were prepared using Li:Fe ratios of 1:1, 1:3 and 1:5 in the form of powders and ceramics. Fig. 1a shows the RBS-NRA spectra measured for the three powders. In the figure, the surface signals from O, Fe and Li are marked by arrows and the Li signal has been magnified on the inset. The measured RBS-NRA spectra show notable differences in terms of the metallic content for the various samples. For instance, <Li-Fe 1–1 powd> seems to have significantly more Li than the other two samples while the Fe signal appears diminished, as expected from the Li:Fe ratios used in the synthesis. Aiming at quantifying these differences in elemental composition, and particularly in the Li:Fe content, relevant RBS-NRA spectra have been simulated using the SIMNRA commercial code [27]. Measured and simulated spectra for a representative sample (<Li-Fe 1–5 powd>) are displayed on Fig. 1b in red and black color, respectively, illustrating the good agreement achieved between experiment and theory.

In Fig. 1b, the independent elemental contributions to the total simulated spectrum are also represented. In addition to the Li, Fe and O present in the sample itself, C was also included in the simulations in order to account for the contribution of the carbon tape on which the powders were fixed for the measurement. The cross-section input related to the lithium nuclear reaction [7] Li(p,α)$^4$He was extracted from Paneta et al. [26] For C and O, the $^{12}$C(p,p)$^{12}$C and $^{16}$O(p,p)$^{16}$O elastic-scattering cross-sections have been used [27]. The Rutherford scattering cross section was used to simulate the Fe intensity.

As a consequence of the samples morphology (powders fixed on carbon tape), the elemental signals of the spectrum show a continuous decrease as the ion beam penetrates deeper in the sample (tails towards lower energies), in contrast with the sharp peaks generally recorded for thin films. The elemental concentrations, determined by SIMRA simulations, were included to fit the spectra. The fit was carried out assuming that the sample is composed by 10 different layers and a porosity of a 40%. Additional fits (not shown here) of the RBS-NRA data have demonstrated the porosity does not significantly affect the calculated elemental composition.

Similar measurements and simulations were carried out for the ceramic samples, producing alike results, although as it appears from the spectra recorded for the pellets, these samples suffered a slight contamination originating from the SiC sandpaper used to grind them to powders for the measurement. Thus, the accuracy of the quantification is lower for the ceramics, as an additional contribution from Si had to be included in the simulation. Table 1 includes the Fe/Li ratios resulting from the simulations for all powder and ceramic samples. The obtained Fe/Li values seem to indicate that the Li:Fe ratios of the starting mixtures are essentially maintained after both the synthesis at 900 °C and sintering at 1100 °C, within the calculated errors. Preparation of Li ferrites through conventional ceramic methods has been traditionally avoided, as the high temperatures required for sintering have been seen to promote the evaporation of Li. [14] However, this has not been the case for the samples under study here, as the Li-content determined on the ceramics is comparable to the Fe/Li ratios of 1, 3 and 5 employed for the synthesis. A possible explanation for stoichiometry preservation in the present samples is the optimization of the ceramic processing carried out here, yielding powders with stoichiometric composition, good crystallinity and particle sizes which allow sintering at a relatively low temperature.

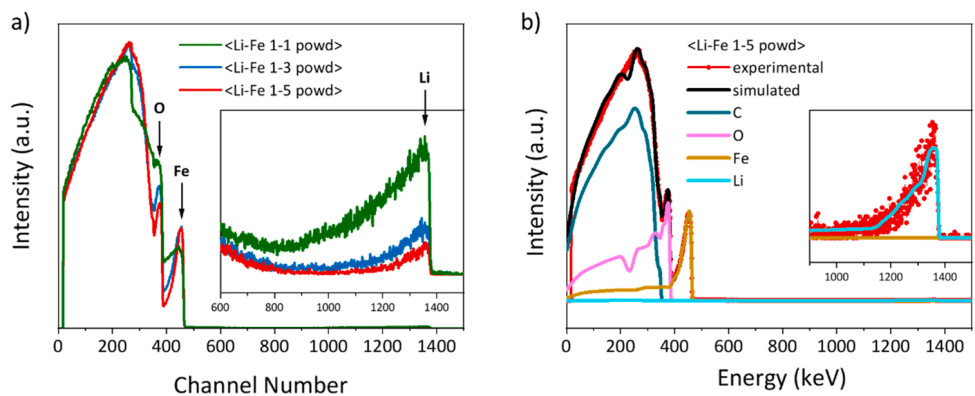

**Fig. 1.** (a) RBS-NRA spectra of the three powder samples. Inset: magnification of the yield corresponding to the lithium signal. (b) Measured and simulated RBS-NRA spectra corresponding to sample <Li-Fe 1–5 powd> in red and black color, respectively, along with the individual elemental contributions to the total simulated spectrum. Inset: magnification of the Li signal.

**Table 1**
Fe/Li ratios obtained from the simulated RBS-NRA spectra and from Rietveld refinements of PXRD data for the different samples.

| sample | Fe/Li ratio from NRA | Fe/Li ratio from PXRD |
|---|---|---|
| <Li-Fe 1–1 powd> | 1.0(1) | 1.12(5) |
| <Li-Fe 1–3 powd> | 3.2(1) | 4.3(2) |
| <Li-Fe 1–5 powd> | 5.2(1) | 4.9(2) |
| <Li-Fe 1–1 ceram> | 0.9(2) | 1.44(6) |
| <Li-Fe 1–3 ceram> | 2.6(2) | 3.9(2) |
| <Li-Fe 1–5 ceram> | 5.1(2) | 5.3(2) |

The measured PXRD patterns were fitted to Rietveld models containing two phases: α-LiFeO$_2$ (rock-salt structure) and α-LiFe$_5$O$_8$ (spinel structure). Fig. 2a shows the fit for a representative sample (see Fig. S2 for remaining samples). Here, the experimental data is represented by the grey symbols while Rietveld models for LiFeO$_2$ and LiFe$_5$O$_8$ are represented by the red and green lines, respectively. The total Rietveld model is the sum of these two individual phases (not shown in the figure) and the discrepancy between the data and the total model is depicted by the black line at the bottom. The bar graph in Fig. 2c illustrates the composition of each sample in mass percentage (wt%). Pure LiFeO$_2$ and LiFe$_5$O$_8$ powders are obtained from Li:Fe ratios of 1:1 and 1:5, respectively. Subsequent sintering of the 1:5 powders do not seem to affect phase composition while a small fraction of LiFe$_5$O$_8$ arises in the 1:1 sample. The 1:3 Li:Fe ratio leads to coexistence of the two phases in the powder sample: LiFe$_5$O$_8$ as the main phase, with 83(1) wt%, and a 16.9 (8) wt% of LiFeO$_2$. After the sintering treatment applied to the 1:3 powders, the minority phase content increases up to 23.4(9) wt%.

The refined cell dimensions for the rock-salt phase are in good agreement with the reported for LiFeO$_2$ (see values in Table S3) [17]. Refined cell parameters for the spinel phase are plotted in Fig. 2b. While all values are within the expected range for LiFe$_5$O$_8$, sample <Li-Fe 1–5 ceram> presents a greater value than all the other samples. This small shift towards larger unit cell dimensions may be understood based on the greater Fe-content of this sample compared to the others (see Table 1 and Table S4), given that the isostructural iron oxide magnetite Fe$_3$O$_4$ is 8.3985(5) Å, [33] although further characterization by means of e.g. Mössbauer spectroscopy is needed to confirm this hypothesis. The large uncertainty on the cell dimensions reported for <Li-Fe 1–1 ceram> is due to the low concentration of LiFe$_5$O$_8$ present in this sample.

As a last stage of the data refinement, atomic occupancies were refined for the metal cations, while the oxygen sites were assumed fully occupied, and from the refinements, the lithium content was calculated. The lithium content on each phase is defined as $x$ in Fig. 3a, $x$ representing the chemical subscript corresponding to Li in the phases defined as Li$_x$Fe$_{1-x}$O$_2$ and Li$_x$Fe$_{5-x}$O$_8$ (check Table S4 for numerical values). In the theoretical LiFeO$_2$ structure, Li$^+$ and Fe$^{3+}$ are equally distributed filling the 4a Wyckoff position (see Table S1 for full phase description). In this work, the occupancies of both atomic positions were refined, revealing that the 4a position has a slight Li-deficiency in all the samples studied (see red symbols in Fig. 3a). The $x$ calculated for the rock-salt phase ranges from 0.84(2) to 0.942(9), slightly below the theoretical value of 1. One could say that the Li-content in LiFeO$_2$ tends to drop as the amount of Fe used in the synthesis increases but the calculated uncertainties do not really allow drawing a clear trend.

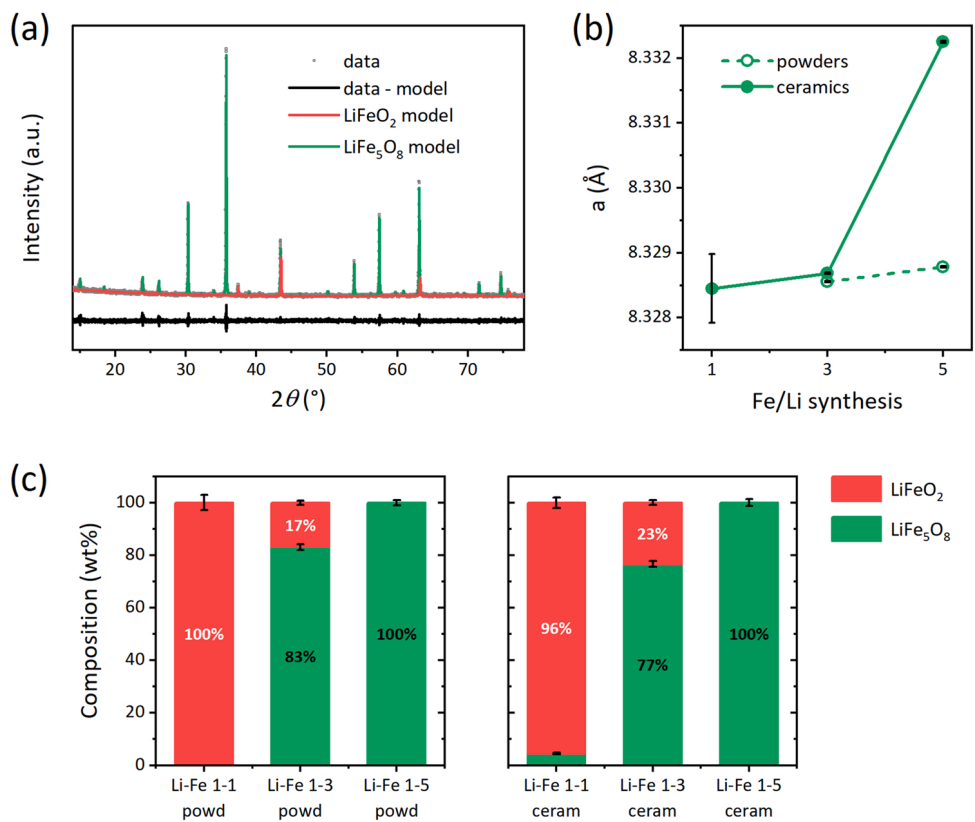

**Fig. 2.** (a) PXRD data and corresponding Rietveld model for sample <Li-Fe 1–3 powd>. (b) Unit cell parameter, a, of the spinel phase and (c) sample composition extracted from Rietveld analysis (left: powders, right: pellets).

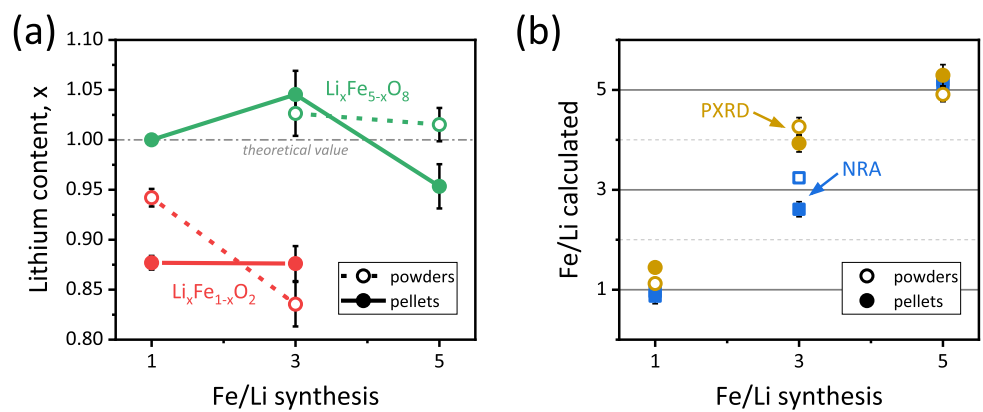

**Fig. 3.** (a) Lithium content on LiFeO$_2$ (red) and LiFe$_5$O$_8$ (green). The theoretical coefficient of Li in both phases is 1. (b) Ratio between the Fe and Li content calculated from Rietveld refinements of PXRD patterns (yellow) and RBS-NRA simulated spectra (blue).

For the LiFe$_5$O$_8$ phase, it has been previously reported that the Li$^+$ cations have a tendency to lodge at the octahedral positions of the structure [13,34,35]. Therefore, in the Rietveld models of the present work, the tetrahedral sites (8c) were always fully occupied by Fe$^{3+}$. On the other hand, the Li$^+$ and Fe$^{3+}$ content on the octahedral sites (with Wyckoff symbols 4a and 12d, respectively) was refined so that the two cations could freely swap with one another as long as both sites remained fully occupied. The obtained trends for the Li-content in the spinel phase are displayed in green color in Fig. 3a. For all samples, the chemical subscript for Li, $x$, is fairly close to the theoretical value of 1 (within a ± 0.05 uncertainty). For sample <Li-Fe 1–1 ceram>, the low significance of the LiFe$_5$O$_8$ phase did not allow for refining atomic occupancies. Similar to the rock-salt scenario, no clear trend is identified for the Li content in the spinel phase, although the obtained results lay in all cases within ± 5% of the theoretical value.

An overall Fe-to-Li content has been calculated considering the refined coefficients and weight fractions for each phase, and they are plotted in Fig. 3b as a function of the Fe/Li ratios used in the synthesis. The calculated ratios correlate well with the nominal compositions, although the calculated values for <Li-Fe 1–3 powd> and <Li-Fe 1–3 ceram> (i.e., 4.3(2) and 3.9(2), respectively) are above the expected value of 3. The Fe/Li values calculated from the RBS-NRA simulations are also plotted in Fig. 3b for comparative purposes. The results from both techniques are in very good agreement for both end compositions, although a discrepancy in the Li-Fe 1–3 samples is noted. It is worth noting that the discrepancy is observed for the samples that have a mixture of phases, while the agreement is better for samples in which one phase predominates.

Averaged Raman spectra of the powders and ceramics on representative superficial regions are displayed in Fig. 4a. The Raman signal corresponding to <Li-Fe 1–3 powd> and <Li-Fe 1–5 powd> are very similar to one another, showing Raman vibrational modes at 128, 202, 237, 264, 301, 322, 259, 282, 403, 440, 493, 521, 553, 611, 666 and 713 cm$^{-1}$. These modes can be associated with the LiFe$_5$O$_8$ spinel phase, and more specifically, with the α polymorph, for which $6A_1 + 14E + 20F_2$ phonon Raman modes are allowed [13,20,22]. In addition, some bands at larger wavenumbers are observed, at around 1161 and 1376 cm$^{-1}$, which are attributed to second-order modes [22]. The spectrum measured for <Li-Fe 1–1 powd> is completely different from the other two powder samples. In this case, three broad vibrational modes are identified at 180, 394 and 624 cm$^{-1}$. This spectrum can be attributed to the LiFeO$_2$ rock-salt phase [1,36,37]. From the group factor analysis, two phonon modes are active in Raman: $A_1 + E_g$. The Raman band at 394 cm$^{-1}$ ($E_g$) is attributed to the O–Fe–O bending while the vibrational mode at 624 cm$^{-1}$ ($A_1$) corresponds to the Fe–O stretching mode of FeO$_6$ octahedra. The vibrational band at around 180 cm$^{-1}$ can be assigned to the Li-cage in an octahedral environment, as identified by C. Julien in Raman spectra for similar materials, i.e, LiMO$_2$ (M= Ni, Co, Cr) [37]. Finally, similar to the spinel structure, a band around 1276 cm$^{-1}$ related to an overtone is also visible.

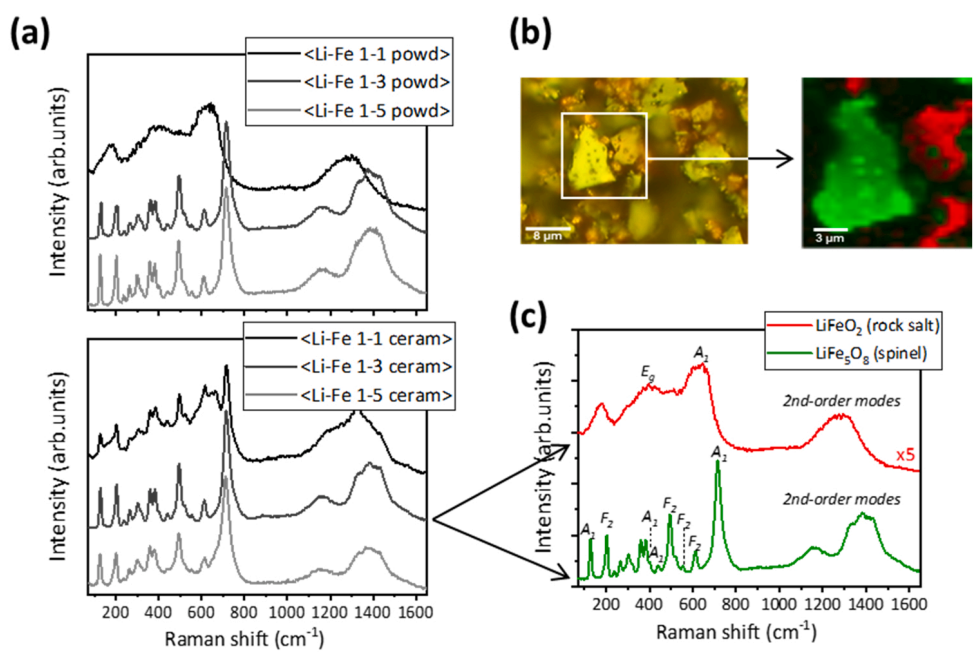

**Fig. 4.** (a) Raman spectra averaged over > 5 × 5 μm$^2$ regions for as-prepared powders and sintered pellets (ceramics). (b) (left) Optical image taken on sample <Li-Fe 1–3 ceram> and (right) in-plane Raman image on the 15 × 15 μm$^2$ region delimited by the white square on the optical image, the colors identify regions of the Raman image with a similar Raman spectrum. (c) Main Raman spectra recognized in the Raman mapping from figure b, corresponding to the rock-salt and spinel phases (in red and green color, respectively). In figure c, the allowed Raman modes for the rock-salt phase are indexed. For the spinel phase, $A_1$ and $F_2$ Raman modes are identified with the corresponding tags while untagged bands correspond to E vibrational modes.

With respect to the sintered pellets fabricated from the powder samples, vibrational modes for <Li-Fe 1–3 ceram> and <Li-Fe 1–5 ceram> are corresponding to the LiFe$_5$O$_8$ phase while <Li-Fe 1–1 ceram> presents a convolution of the two identified Raman spectra: LiFe$_5$O$_8$ + LiFeO$_2$. Subtle changes in the position, full width high maximum (FWHM) and intensity of the Raman bands are noted when comparing powder and ceramic samples. The fluctuations in the relative intensities of the bands can be attributed to different crystallographic orientations of the grains, while the subtle differences in position and FWHM may be associated with slight changes on the grain sizes or small strain that may be induced in the structure during the synthesis or sintering processes. No binary Li or Fe oxides are identified in any sample [38,39].

With the aim of gaining semi-quantitative information on the concentration and distribution of each of the phases, spatially-resolved analysis of the confocal Raman data has been carried out. Fig. 4b shows the corresponding analysis for sample <Li-Fe 1–3 ceram>. Fig. 4b (right) shows the in-plane Raman image corresponding to the region marked on the optical image from the left. In the 2D Raman image, two distinct phases are discriminated, with regions colored in red and green corresponding to the rock-salt and spinel phases, respectively. From both the Raman and optical images, it is roughly estimated that the rock-salt phase accounts for about 1/3 of the total surface mapped, which is in rather good agreement with the 23.4(9) wt% extracted from Rietveld analysis for that phase in sample <Li-Fe 1–3 ceram>. Raman map analysis for the remaining the samples may be found in Fig. S4, yielding results compatible with those from Rietveld analysis.

Average Raman spectra corresponding to the two colored regions from Fig. 4b are plotted in Fig. 4c in matching colors. In Fig. 4c, the rock-salt Raman spectrum (red) has been multiplied by a factor 5 to achieve an overall intensity comparable to that of the spinel one (green). The need for this x5 factor highlights the great difference in Raman cross section of these two phases. As a consequence of this imbalance between phases, the average Raman spectra for the samples in this work do not give a good idea of the actual concentration of each phase. This explains why the average Raman spectrum for <Li-Fe 1–3 ceram>, plotted at the bottom of Fig. 4a, gives the impression of a nearly pure LiFe$_5$O$_8$ sample, despite being calculated from the superficial region shown in Fig. 4b, with a clear contribution from LiFeO$_2$. The same happens for the corresponding powder sample Li-Fe 1–3 powd>. This is even more noticeable for <Li-Fe 1–1 ceram>, which according to PXRD is nearly single-phase LiFeO$_2$ (96(2) wt%) while the averaged Raman spectra shown in Fig. 4a seems to be predominantly spinel phase. Spatially-resolved analysis of the data removes some of the uncertainties that might derive from phase identification based on average Raman spectra exclusively, evidencing the need for Raman mapping experiments.

### 3.2. Particle size distribution, morphology and densification

Fig. 5a-c show FE-SEM micrographs for the three powder samples. In all three cases, relatively loose particles with fairly isotropic morphology are observed. Particle sizes extracted from the micrographs were successfully fitted to a lognormal distribution, yielding similar mean values of ≈ 600 nm for all three powders (within uncertainties, see Table 2). Size distributions differ to some extent, with <Li-Fe 1–1 powd> showing a wider distribution with somewhat more representation of large sizes than <Li-Fe 1–3 powd>, and the latter also more than <Li-Fe 1–5 powd>.

The microstructure of fresh fractured surfaces of the ceramics is shown in Fig. 5d-f. For <Li-Fe 1–1 ceram> (Fig. 5d), the former particle morphology has evolved to grains with irregular morphology and faceted grain boundaries which are interconnected through sintering necks. The grains have experimented some growth, reaching a mean value of 3.8(6) μm, which proves insufficient to eliminate the interconnected porosity observed in the sample. The fracture is mainly intergranular, suggesting that mass transport during sintering is limited to coalescence of clusters of particles. All these evidences are indicative of an early sintering stage, which is also in agreement with the temperature behavior displayed by the powders employed to fabricate this ceramic. As shown by the shrinkage curve for <Li-Fe 1–1 powd> extracted from hot-stage microscopy measurements (see Fig. S5a), at the sintering temperature used (i.e., 1100 °C), mass transport is rather limited and the compound has barely started shrinking, < 8%. The sintering process has reached a more advanced stage for <Li-Fe 1–3 ceram> (Fig. 5e), with appearance of grains with large sizes, although the lognormal mean value of 3.4(2) μm is, within uncertainties, equivalent to that for the 1–1 ceramic (see Table 2). The microstructure of <Li-Fe 1–3 ceram> is a combination of the observed for the 1–1 and 1–5 ceramics –not surprising given the mixed composition of this intermediate sample– and a mixture of both intra- and intergranular fractured surfaces are found. <Li-Fe 1–5 ceram> (Fig. 5f) has come to a much further sintering stage than the other two ceramics. Characteristic 3-point junctions with 120° angles appear for <Li-Fe 1–5 ceram>, indicating the sintering process approaches its final stage [40]. In the 1–5 ceramic, the grains have grown substantially –finding particles over 30 μm in diameter– and size distribution expands considerably. A mean grain size of 10(2) μm is extracted for the lognormal fit. The fractured surface of <Li-Fe 1–5 ceram> is clearly intragranular, as expected when grain growth occurs during sintering. All this is again understood based on the shrinkage curve of <Li-Fe 1–5 powd> (see Fig. S5b), which evidences that for this composition, the greatest mass transport (and in turn, the maximum densification speed) takes place at a temperature near the sintering point (i.e., 1100 °C).

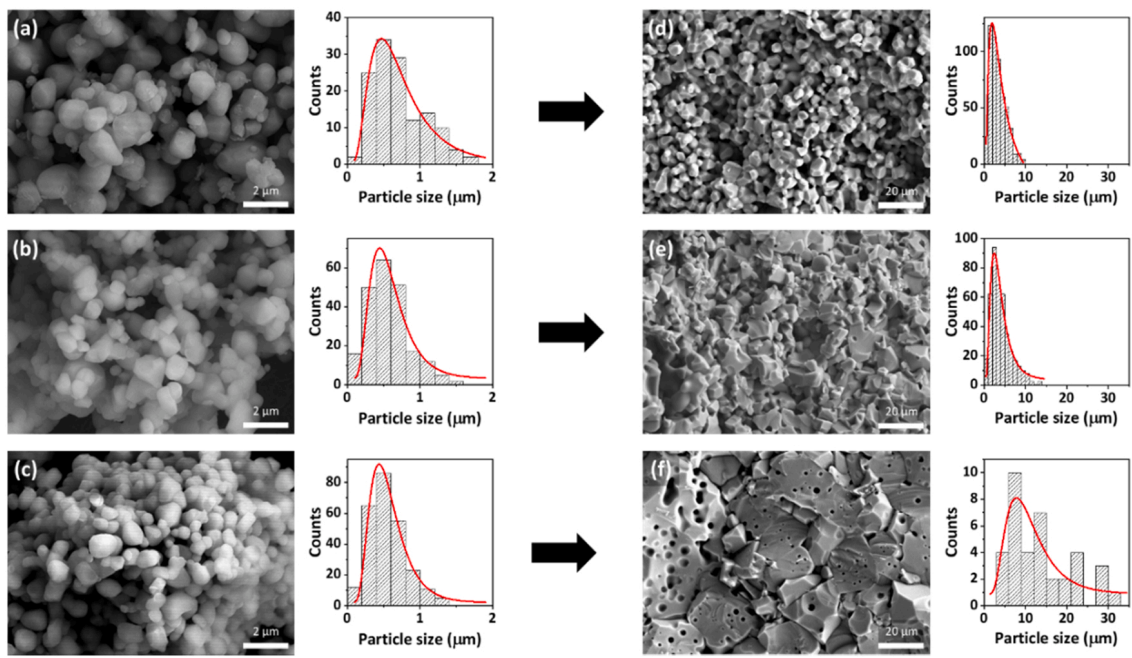

**Fig. 5.** FE-SEM micrographs for (a) <Li-Fe 1–1 powd>, (b) <Li-Fe 1–3 powd>, (c) <Li-Fe 1–5 powd>, (d) <Li-Fe 1–1 ceram>, (e) <Li-Fe 1–3 ceram> and (f) <Li-Fe 1–5 ceram>. Particle size distribution based on the relevant image, data fitted to a lognormal distribution.

**Table 2**
Lognormal mean particle/grain sizes obtained from FE-SEM. Li-O bond distances obtained from Rietveld refinements of PXRD data. Absolute and relative densities measured for the pellets.

| sample | Particle (powders) or grain (ceramics) mean size (μm) | Li-O on rocksalt (Å) | Li-O on spinel (Å) | Archimedes density (g/cm3) | Relative density (%) |
|---|---|---|---|---|---|
| <Li-Fe 1–1 powd> | 0.64(6) | 2.07909 | - | - | - |
| <Li-Fe 1–3 powd> | 0.54(4) | 2.07830 | 2.11292 | - | - |
| <Li-Fe 1–5 powd> | 0.53(2) | - | 2.11297 | - | - |
| <Li-Fe 1–1 ceram> | 3.8(6) | 2.07847 | 2.11290 | 3.26 | 74.3(3) |
| <Li-Fe 1–3 ceram> | 3.4(2) | 2.07822 | 2.11295 | 4.07 | 87.1(5) |
| <Li-Fe 1–5 ceram> | 10(2) | - | 2.11385 | 4.41 | 92.7(6) |

A pronounced grain growth generally comes as a consequence of a fast grain boundary mobility. As grain boundary mobility is typically faster than the processes required to remove the air between particles, it is common for pores to appear in these conditions. For <Li-Fe 1–5 ceram>, a great number of trapped pores are visible and different stages of pore coalescence are found (see Fig. 5f). Similar pores have been previously observed on sintered LiFe$_5$O$_8$ samples [41–43]. Coalescence of intragranular pores and subsequent pore removal involves displacement of the air trapped in the pores towards the surface, and consequently, it requires much more energy than grain boundary diffusion. Hence, the rapid grain growth observed here suggests that, in this system, mass transport during sintering is assisted by grain boundary diffusion, while pore coalescence may occur only to a small extent. Li has a high vapor pressure at the selected sintering temperature and therefore, one could expect vapor transport processes to be favored for these materials [44]. However, the preservation of the Li-content after sintering denotes that vapor-phase assisted processes must not be dominant at the selected sintering temperature, although small contributions from such processes to the mass transport cannot be completely ruled out.

While the presence of intragranular pores is especially relevant for <Li-Fe 1–5 ceram>, a few of these pores are also visible in <Li-Fe 1–3 ceram> while no pores are found in <Li-Fe 1–1 ceram>. This together with the grain growth observed for each sample, suggests that mass transport diffusion must be much more favorable in LiFe$_5$O$_8$ than in LiFeO$_2$, despite the fact that the latter has a higher proportion of Li. This is counterintuitive a priori, given that Li-cations are known for their high mobility, which would in principle be expected to favor mass transport. A possible explanation of this phenomenon could reside on the crystal structures of these two materials. In both rock-salt and spinel structures, Li$^+$ cations occupy the octahedral sites defined by six O$^{2-}$ anions. The Li-O bond distances extracted from Rietveld analysis of PXRD data (see Table 2) are shorter for LiFeO$_2$ than for LiFe$_5$O$_8$, meaning that Li-cations are less strongly bound to the oxygen anions in the spinel structure, which might be the reason why diffusion is greater in LiFe$_5$O$_8$, in spite of its lower Li-content.

The relative densities measured for the ceramic samples (see Table 2) are in good agreement with the sintering stage attained for each. Thus, the density is greater for <Li-Fe 1–5 ceram> than for <Li-Fe 1–3 ceram>, while the lowest value is measured for <Li-Fe 1–1 ceram>. In all three cases, the attained densities ensure a good mechanical integrity, which together with the Li:Fe stoichiometry preservation demonstrated, makes them suitable for many applications, including their use as targets for thin-film deposition by means of e.g. sputter or pulsed-laser deposition (PLD). This is of great use considering many of the applications of Li ferrites require a thin-film morphology. Density improvement is generally expected from increasing the sintering temperature, however, a higher working temperature would increase the Li-cation partial pressure, very likely altering the Li:Fe ratio of the produced ceramic. Based on the sintering mechanisms inferred from the FE-SEM microstructural analysis, it follows that density betterment should come from strategies promoting mass transport through surface diffusion mechanisms while retaining the Li-cation partial pressure as low as possible. In particular, the use of additives on the surface of the particles of alike systems has been previously proposed as a suitable route to reduce the sintering temperature and allow an effective control of grain growth [41,45,46]. However, the present work demonstrates that that optimization of the ceramic processing also allows producing dense ceramics with stoichiometry preservation, which is key for their technological application.

### 3.3. Magnetic properties

Magnetization, $M$, as a function of an externally applied magnetic field, $H_{app}$, has been measured at RT on the ceramic samples and the recorded hysteresis loops are plotted in Fig. 6. <Li-Fe 1–5 ceram> exhibits a low-coercivity cycle, as expected for a soft ferrimagnet (FiM) such as LiFe$_5$O$_8$. Both Raman microscopy and powder diffraction agree upon this sample being phase-pure LiFe$_5$O$_8$, and the saturation magnetization, $M_s$, of 61.5(1) Am$^2$/kg measured here fits well into this scenario – the $M_s$ values reported for dense LiFe$_5$O$_8$ pieces span within 57–63 Am$^2$/kg [23,47,48]. From CRM and PXRD it follows that LiFeO$_2$ is the majority phase in <Li-Fe 1–1 ceram>. LiFeO$_2$ is an antiferromagnet (AFM) with an ordering temperature below RT, and therefore, no hysteretic behavior is to be expected for this material at RT. A small hysteresis loop is recorded for <Li-Fe 1–1 ceram> (see red curves in Fig. 6), which can only arise from the little LiFe$_5$O$_8$ present in the sample (4.4(3) wt% according to Rietveld analysis). In this case, the $M_s$ value is almost null, given the predominance of the paramagnetic (PM) phase and low concentration of FiM material. <Li-Fe 1–3 ceram> is a mixture of the two magnetic materials described above (77(1) wt% LiFe$_5$O$_8$, 23.4 (9) wt% LiFeO$_2$ according to diffraction). It presents a hysteresis curve similar to that of <Li-Fe 1–5 ceram>, only the $M_s$ decreases down to 47.2(1) Am$^2$/kg. Phase composition (wt% LiFe$_5$O$_8$) for each sample can be roughly estimated from the measured $M_s$ with the formula $M_s$(sample) = $M_s$(⟨Li − Fe 1 − 5 ceram⟩) × wt% $LiFe_5O_8$, assuming that (i) <Li-Fe 1–5 ceram> is phase-pure $LiFe_5O_8$ and (ii) $M_s$ for the PM phase should be ≈ 0. The results from these calculations –collected in Table 3, along with the $M_s$ values– are in very good agreement with the wt% $LiFe_5O_8$ obtained from Rietveld analysis of PXRD data (see Fig. 2c).

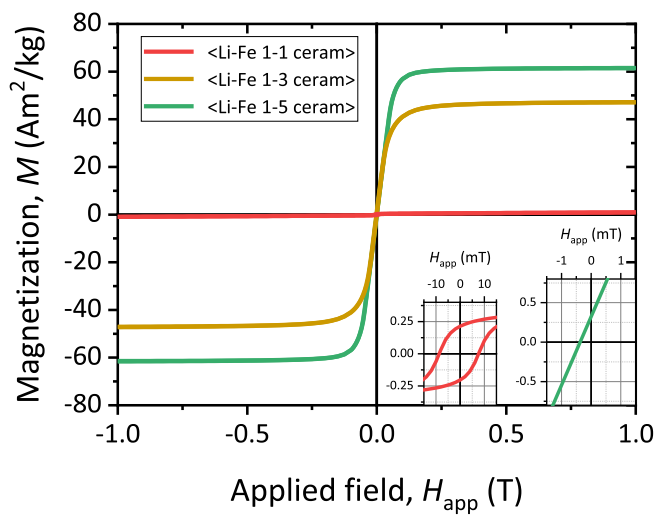

Fig. 6. RT magnetization (M) versus applied field ($H_{app}$) hysteresis loops for the dense ceramics. On the insets, the near H = 0 region is magnified for <Li-Fe 1–1 ceram> (red) and <Li-Fe 1–5 ceram> (green) to qualitatively demonstrate the behavior of $H_c$.

Table 3
Saturation magnetization, $M_s$, and coercivity, $H_c$, extracted from measured hysteresis loops and wt% $LiFe_5O_8$ calculated from the $M_s$ values, assuming <Li-Fe 1–5 ceram> is phase-pure $LiFe_5O_8$.

| sample | $M_s$ (Am$^2$/kg) | $H_c$ (mT) | wt% |
|---|---|---|---|
| <Li-Fe 1–1 ceram> | 0.96(6) | 8(2) | 2% |
| <Li-Fe 1–3 ceram> | 47.2(1) | < 4 | 77% |
| <Li-Fe 1–5 ceram> | 61.5(1) | < 4 | 100% |

Unfortunately, the coercivity, $H_c$, values extracted from the curves are near the resolution limit of the measurement. During the measurement, the $H_{app}$ was swept in steps of 2 mT, so this is considered as the measurement error. Therefore, we cannot rigorously determine any $H_c$ < 4 mT, although a clear difference between <Li-Fe 1–1 ceram> and <Li-Fe 1–5 ceram> is qualitatively observed. Note that even though the $H_c$ in <Li-Fe 1–1 ceram> is a consequence of the $LiFe_5O_8$ present in the sample, the $H_c$ measured here is much higher than that for the phase-pure material. Such magnetic hardening (increase in $H_c$) can be explained if picturing the sample as a set of small deposits of FiM material ($LiFe_5O_8$) dispersed through a non-magnetic matrix ($LiFeO_2$). The dilution of ferri- or ferromagnetic (FiM, FM) particles reduces the dipolar interaction among them, causing an enhancement of $H_c$; this effect has been previously reported and is well understood [49,50]. On the other hand, the coercivity value measured for <Li-Fe 1–5 ceram> is much lower (see Fig. 6 inset), which is believed to be closely related to the severe grain growth this sample undergoes as a consequence of the sintering treatment at 1100 °C, clearly yielding the FiM material in a multi-domain magnetic configuration. Besides the pores mentioned earlier, the grains in <Li-Fe 1–5 ceram> seem fairly continuous and homogeneous (see FE-SEM in Fig. 5f), as also demonstrated by the rather high density of the piece (relative density of 92.7(6)%). This microstructure heavily favors the motion of the magnetic domain walls throughout the large grains, which is known to be the most probable mechanism for magnetization reversal in bulk samples [12,51]. Low-hampered domain wall motion implies a rapid propagation of any disturbance caused by an external field, therefore diminishing the coercive field of the material. Previously reported $H_c$ values for dense $LiFe_5O_8$ pellets are considerably scattered, ranging between 0.4 and 16 mT [23,24,35,48,52], and similar values are found for powders [21,53,54].

## 4. Conclusions

Lithium ferrite samples have been prepared as loose powders and sintered ceramics, attaining a relative density of 92.7(6)% for a phase-pure $LiFe_5O_8$ pellet. The lithium content has been evaluated based on Rutherford backscattering spectroscopy combined with nuclear reaction analysis and Rietveld analysis of powder X-ray diffraction data, in both cases confirming that the Li:Fe ratios remain relatively steady during the synthesis at 900 °C and sintering at 1100 °C, yielding a final dense product with controlled lithium content. This is in contrast with previous works reporting that the high temperatures required by conventional ceramic methods promote lithium evaporation and yield Li-deficient products. Preservation of the Li:Fe stoichiometry of the ceramics prepared here is attributed to a combination of both good quality of the synthesized powders and the relatively low sintering temperature employed. The controlled stoichiometry and mechanical integrity of the ceramics prepared here allows their direct utilization in bulk shape, in addition to their use as targets for thin-films deposition. Field emission scanning electron microscopy has revealed the substantial grain growth taking place during sintering, going from ≈ 600 nm powders to $LiFeO_2$ and $LiFe_5O_8$ ceramics with mean sizes of 3.8(6) and 10(2) μm, respectively. Large pores have been detected on the ceramics, especially on the phase-pure $LiFe_5O_8$ material, which have been attributed to an increased grain boundary mobility by surface diffusion, also compatible with the intense grain growth observed. Additionally, the increased mobility and rapid growth has been seen to be more favored for $LiFe_5O_8$ than for $LiFeO_2$, despite the lower Li-content of the first. The phase-pure $LiFe_5O_8$ ceramic shows a soft ferrimagnetic hysteretic behavior with a saturation magnetization of 61.5(1) Am$^2$/kg while the pellet with Li:Fe = 1:1 displays the antiferromagnetic behavior expected for $LiFeO_2$. Phase compositions estimated from the measured $M_s$ are in good agreement with that obtained from Rietveld analysis of PXRD data.

**Declaration of Competing Interest**

The authors declare that they have no known competing financial interests or personal relationships that could have appeared to influence the work reported in this paper.

**Acknowledgements**

This work has been supported by grants RTI2018-095303-B-C51 and RTI2018-095303-A-C52 funded by MCIN/AEI/ 10.13039/501100011033 and by "ERDF A way of making Europe" and grants PID20210124585NB-C31, PID2021–124585NB-C32 and PID2021-124585NB-C33 funded by MCIN/AEI/ 10.13039/501100011033 and by the "European Union NextGenerationEU/PRTR". C.G.-M. acknowledges financial support from grant FJC2018–035532-I funded by MCIN/AEI/ 10.13039/501100011033 and grant RYC2021–031181-I funded by MCIN/AEI/10.13039/501100011033 and by the "European Union NextGenerationEU/PRTR". A.S. acknowledges financial support from the Comunidad de Madrid for an "Atracción de Talento Investigador" contract No. 2017-t2/IND5395 and grant RYC2021-031236-I funded by MCIN/AEI/10.13039/501100011033 and by the "European Union NextGenerationEU/PRTR". A.Q. acknowledges financial support from grant RYC-2017023320 funded by MCIN/AEI/ 10.13039/501100011033 and by "ESF Investing in your future". The authors acknowledge support from CMAM for beamtime proposals with codes STD019/20, STD026/20 and STD033/20.

**Appendix A. Supporting information**

Supplementary data associated with this article can be found in the online version at doi:10.1016/j.jeurceramsoc.2023.02.011.